\documentclass[12pt]{article}
\usepackage{amssymb}
\begin{document}
\baselineskip=18pt
\pagenumbering{arabic}
\parskip1.5em
\newcommand{\ee}{{{\eta}\over {2}}}
\newcommand{\beq}{\begin{equation}}
\newcommand{\eeq}{\end{equation}}
\newcommand{\beqa}{\begin{eqnarray}}
\newcommand{\eeqa}{\end{eqnarray}}
\newcommand{\beqan}{\begin{eqnarray*}}
\newcommand{\eeqan}{\end{eqnarray*}}
\newcommand{\half}{{{1}\over{2}}}
\newcommand{\ihalf}{{{i}\over{2}}}
\newcommand{\quar}{{{1}\over{4}}}
\newcommand{\la}{\lambda}
\newcommand{\si}{\sigma}
\newcommand{\Si}{\Sigma}
\newcommand{\tf}{\Theta}
\newcommand{\3}{{\ss}}
\newcommand{\bra}{\langle 0|}
\newcommand{\ket}{|0\rangle}
\newcommand{\id}{{1}\hspace{-0.3em}{\rm{I}}}
\newcommand{\tn}[1]{T^{#1}}
\newcommand{\tr}{\bigtriangleup}
\newcommand{\trb}{\bar{\bigtriangleup}}
\def\b{\beta}
\def\d{\delta}
\def\g{\gamma}
\def\a{\alpha}
\def\s{\sigma}
\def\t{\tau}
\def\l{\lambda}
\def\e{\epsilon}
\def\r{\rho}
\def\d{\delta}
\def\wid{\widehat}
\def\ds{\displaystyle}
\def\be{\begin{equation}}
\def\ee{\end{equation}}
\def\beq{\begin{eqnarray}}
\def\eeq{\end{eqnarray}}
\def\ov{\overline}
\def\om{\omega}
\thispagestyle{empty}
\begin{flushright}BN--TH--2000--01\\
\end{flushright}
\vskip2.5em
\begin{center}
{\Large{\bf Resolution of the Nested Hierarchy for Rational sl(n) Models}}\vskip1.5em
T.-D. Albert\hspace{3em}H. Boos$^{\natural}$\hspace{3em}R. Flume\hspace{3em}K. Ruhlig\\\vskip3em
{\sl Physikalisches Institut der Universit\"at Bonn}\\
{\sl Nu{\ss}allee 12}\\
{\sl D--53115 Bonn}\\
{\sl Germany}
\end{center}
\vskip2em
\begin{abstract}
\noindent We construct Drinfel'd twists for the rational $sl(n)$  XXX-model giving rise to a completely symmetric representation of the monodromy matrix.\\
We obtain a polarization free representation of the pseudoparticle creation operators figuring in the construction of the Bethe vectors within the framework of the quantum inverse scattering method. This representation enables us to resolve the hierarchy of the nested Bethe ansatz for the $sl(n)$ invariant rational Heisenberg model. Our results generalize the findings of Maillet and Sanchez de Santos for $sl(2)$ models.
\end{abstract}\vskip5em
$^{\natural}$ {\small{on leave of absence from the Institute of High Energy Physics, Protvino, Russia}}\\
{\small{
e-mail: t-albert@th.physik.uni-bonn.de\\
\hspace*{1.3cm}boos@th.physik.uni-bonn.de\\
\hspace*{1.3cm}flume@th.physik.uni-bonn.de\\
\hspace*{1.3cm}ruhlig@th.physik.uni-bonn.de}}
\vfill\eject\setcounter{page}{1}

\section{Introduction}

In a seminal  paper  Maillet  and  Sanchez de Santos  \cite{ms}   revealed  
the uses of factorizing Drinfel'd twists \cite{drin} for inhomogeneous statistical 
spin chain models for which the method of the algebraic Bethe ansatz is available. 
Those authors used as paradigmata  of their argumentation  the rational XXX  and 
the trigonometric XXZ models  being realized  on tensor products of two-dimensional 
(fundamental) representations of the underlying group $sl(2)$. They showed that the 
similarity transformation provided by the Drinfel'd twist gives rise to a completely 
symmetric representation  of the respective monodromy matrices and implies  simplifying 
features in the new basis - to be described in detail below - for the various operators in the grid of the 
monodromy matrix.\\
The results of  \cite{ms} have been  generalized to any finite dimensional irreducible 
representation of the Yangian $Y(sl(2))$ \cite{terras} and have been used to achieve 
substantial simplifications in the calculation of form factors \cite{maillet}, in 
the determination of thermodynamic quantities such as the spontaneous magnetization 
\cite{iz}, and to solve the so called quantum inverse problem \cite{qis1}, \cite{qis2}, 
that is, to express the local spin operators of the microscopic model through the 
operators figuring in the algebraic  Bethe ansatz.\\
The most striking aspect of the results in \cite{ms} is, as we think, related to the 
fact that  no polarization clouds are attached to quasiparticle creation and annihilation 
operators in the basis in which the monodromy matrix is completely symmetric. This means 
in terms of a particle notation that no virtual particle--antiparticle pairs are present 
in the wave vectors generated by the action of the creation operators to the ground 
state (the reference state of the Bethe ansatz), or in spin chain terminology that 
the creation  and annihilation  operators  are exclusively  built from local spin 
raising and spin lowering operators respectively (that is,  there are no compensating 
pairs of local raising and lowering spin operators). It was noted in \cite{ms}  that 
this latter feature underscores the neat connection between the quantum spin chain 
models and their respective quasiclassical limits, which are  Gaudin magnets \cite{gaudin}, insofar as the appearance of the quasi particle operators of the quantum models 
in the particular basis  differ from the corresponding operators in the quasiclassical 
limit models  only by a ``diagonal dressing'' (see below).\\
This connection motivated us to attempt a generalization of the work of Maillet and 
Sanchez de Santos towards models based on higher rank groups. We will deal here with 
the simplest conceivable extension in the form of the rational XXX model with $sl(n)$ 
as underlying group.\\
A notorious technical difficulty of integrable models with underlying higher rank  
group arises from the intricacies of the recursive procedure of the hierarchical 
Bethe ansatz \cite{kulresh}. It has been known for some time \cite{flume} that the 
recursion of the hierarchical ansatz can be resolved in the case of the quasiclassical 
limit of the rational models, i.e., the rational Gaudin magnets. Constructing the 
analogue of the factorizing twist of \cite{ms} for higher rank models one may hope - 
in view of the affinity of the special basis rendered  by the factorizing twist with 
the quasiclassical limit model - for an explicit resolution of the Bethe ansatz hierarchy. 
This will  indeed be our main result for the spin model under consideration: an 
explicit representation of the $sl(n)$ Bethe wave vectors, solving therewith 
(for the wave vectors) the hierarchy.\\
The plan of  the paper is as follows: section 2 sets the notation, section 3 
is devoted to the construction of the factorizing twist. In section 4 we give 
the expressions  for the $sl(n)$ generators and for the operators contained in 
the monodromy matrix in the basis mediated by the factorizing twist. In section 
5 we discuss the resolution of the Bethe hierarchy. Section 6 contains our 
conclusions. Some technical details are relegated to appendices.

\section{Basic definitions and notation}
Below we shall use many of the notations of references \cite{ms}, \cite{maillet}. We consider the $sl(n)$ Yangian $R$-matrix depending on a spectral parameter $\l$ and a quantum deformation parameter $\eta$:
\be
R_{12}(\l) = b(\l)\id_{12}+ c(\l)P_{12}
\label{R}
\ee
where 
\be
b(\l)={\l\over{\l+\eta}},\quad c(\l)={\eta\over{\l+\eta}}\,.
\label{bc}
\ee
The matrix $R_{12}$ is meant to represent a map $\mathbb{C}^n_{(1)}\otimes\mathbb{C}^n_{(2)}\,\rightarrow\,\mathbb{C}^n_{(1)}\otimes\mathbb{C}^n_{(2)}\;(\mathbb{C}^n_{(1)}\cong\mathbb{C}^n_{(2)}\cong\mathbb{C}^n)$ and $P_{12}$ is the permutation operator acting in $\mathbb{C}^n_{(1)}\otimes\mathbb{C}^n_{(2)}$.
Local spectral parameters attached to vectorspaces $\mathbb{C}_{(i)}^{n}$ isomorphic to $\mathbb{C}^n$ will be called $z_i$. We will also use the notation 
\beq
b_{ij}=b(z_i-z_j), \quad c_{ij}=c(z_i-z_j).
\label{bc1}
\eeq
It is well known that $R$-matrices defined by (\ref{R}) satisfy the
Yang-Baxter equation in vertex form:
\be
R_{12}(z_1-z_2)R_{13}(z_1-z_3)R_{23}(z_2-z_3)=R_{23}(z_2-z_3)R_{13}(z_1-z_3)R_{12}(z_1-z_2)
\label{YB}
\ee
and the unitarity relation
\be
R_{12}R_{21}=\id.
\label{inv}
\ee
where $R_{ij}=R_{ij}(z_i-z_j)$ acts non-trivially on the tensor product $\mathbb{C}_{(i)}^n\otimes\mathbb{C}_{(j)}^n$.\\
Our convention for the matrix indices is as follows:
\beqa 
\left(Z\right)^{\g\,\d}_{\b\,\a}=\left(XY\right)^{\g\,\d}_{\b\,\a}=\left(X\right)^{\g\,\d}_{j_1\,j_2}\left(Y\right)^{j_1\,j_2}_{\b\,\a}\,.\label{matrixproduct}
\eeqa
With the notation $T_{0,23}=R_{03}R_{02},\;\;R_{0i}\equiv R_{0i}(z_i)$,  where the index $0$ refers to an  auxilliary 
space $\mathbb{C}^n_{(0)}$, one may rewrite Eq. (\ref{YB}) in the form of a Faddeev--Zamolodchikov relation
\be
R^{\s_{23}}_{23}T_{0,23}=T_{0,32}R^{\s_{23}}_{23}  \label{lcr}
\ee
with $\s_{23}$ the transposition of space labels $(2,3)$.\\
We use here and subsequently a notation (which may not be in line with common use) that the labels in the upper row are permuted relative to lower indices according to the permutation inscribed, which reads in the example at hand as
$\left(R^{\s_{23}}\right)_{\b_3\,\b_2}^{\a_2\,\a_3}$.\\
It is straightforward to generalize Eq. (\ref{lcr}) to a N-fold tensor product of spaces:\\
With the definition $T_{0,1\ldots N}=R_{0N}\ldots R_{01}$ the generalization reads
\be
R^{\sigma}_{1\ldots N}T_{0,1\ldots N}=T_{0,\sigma(1)\ldots \sigma(N)}R^{\sigma}_{1\ldots N} \label{lcrperm}
\ee
where $\sigma$ is now an element of the symmetric group $S_N$ and $R^{\sigma}_{1\ldots N}$ 
denotes a product of $R$-matrices occuring in (\ref{lcr}), the product corresponding to a decomposition 
of $\sigma$ into elementary transpositions. \\
The order of the upper matrix indices $\a_i$ of the $R^{\s}$ reads according to the above prescription as follows:
\beqa
\left(R^{\s}_{1\ldots N}\right)^{\a_{\s(N)}\ldots\a_{\s(1)}}_{\quad\b_{N}\ldots\b_{1}}\,.\label{Rconvention} 
\eeqa
Eq. (\ref{lcrperm}) implies the composition law (note the difference to the composition law used in ref. \cite{ms})
\beq
R^{\sigma'\sigma}_{1\ldots N}=R^{\sigma}_{\sigma'(1)\ldots \sigma'(N)}  R^{\sigma'}_{1\ldots N}\label{complaw} 
\eeq
for a product of two elements in $S_N$. The factor $R^{\sigma'}_{\sigma(1)\ldots \sigma(N)}$ on the r.h.s.of Eq. (\ref{complaw}) satisfies for itself the relation
\beq
R^{\sigma'}_{\sigma(1)\ldots \sigma(N)}T_{0,\sigma(1)\ldots \sigma(N)}=T_{0,\sigma\sigma'(1)\ldots \sigma\sigma'(N)}R^{\sigma'}_{\sigma(1)\ldots \sigma(N)}.
\eeq

\section{The $F$-matrix and some of its properties}

The starting point of paper \cite{ms} is the Drinfel'd factorizing twists of
the elementary $sl(2)$ $R$-matrix:
$$
R_{12}=F^{-1}_{21}F_{12}
$$
where $F_{12}$ is given by the formula (90) of \cite{ms}                        
\be
F_{12}\; =\; \left(\matrix{
1&{\;0\;}&{\;0\;}&{\;0\;}\cr
0&{\;1\;}&{\;0\;}&{\;0\;}\cr
0&c(z_1-z_2)&b(z_1-z_2)&0\cr
0&{\;0\;}&{\;0\;}&{\;1\;}\cr}\right).
\label{F12}
\ee
The generalization of this formula to the $sl(n)$ case is
of the form
\be
F_{12}=\sum_{n\ge\a_2\ge\a_1}P_{\a_1}^{1}P_{\a_2}^{2}\,\id_{12}+\sum_{n\ge\a_1>\a_2}P_{\a_1}^{1}P_{\a_2}^{2}
R_{12}^{\s_{12}}\,.
\label{F12a}
\ee
Here ${[P^{i}_{\a}]}_{k,l}=\d_{k,\a}\d_{l,\a}$ is the projector on the $\a$
component acting in $i$-th space.\\
Generalizing this factorization matrix to the $N$-site problem one has to satisfy 
at least three properties for the $F$-matrix (see \cite{ms},\cite{maillet}):

\begin{description}
\item[A.]$\quad$ factorization, that is  
\be
F_{\s(1)\ldots\s(N)}(z_{\s(1)},\ldots,z_{\s(N)}) 
R^{\s}_{1\ldots N}(z_{1},\ldots,z_{N}) = F_{1\ldots N}(z_{1},\ldots,z_{N})
\label{factor}
\ee
for any permutation  $\s\in S_N$;

\item[B.]$\quad$ lower-triangularity;

\item[C.]$\quad$ non-degeneracy.
\end{description}

{\bf Proposition 3.1}  The following expression for the $F$-matrix:
\be
F_{1\ldots N} = \sum_{\s\in S_N}\sum^{\quad\quad *}_{\a_{\s(1)}\ldots\a_{\s(N)}}
\prod_{i=1}^N P^{\s(i)}_{\a_{\s(i)}} R_{1\ldots N}^{\s}(z_{1},\ldots,z_{N})
\label{F}
\ee
satisfies the properties {\bf A},{\bf B} and {\bf C}.
The sum $\sum^{*}$ is to be taken over all 
non-decreasing sequences of the labels $\a_{\s(i)}$ which are increasing at
places where the permuted index is decreasing ($\s(i+1)<\s(i)$), namely,
labels $\a_{i}$ should satisfy one of two inequalities for each pair of neighbouring  spaces labels:
\beq
&\a_{\s(i+1)}\ge\a_{\s(i)}\quad \mbox{if}\quad\s(i+1)>\s(i)&\nonumber\\
&\a_{\s(i+1)}>\a_{\s(i)}\quad \mbox{if}\quad\s(i+1)<\s(i)&.
\label{cond}
\eeq

{\bf Proof} $\quad$ First of all let us note that the lower-triangularity can be traced
back to the form of the elementary $R$-matrix using the definition of $F$, Eq. (\ref{F}). Indeed, the ordering (\ref{cond})
just corresponds to the lower-triangularity of the matrix $F$.
Non-degeneracy follows from the lower-triangularity and the fact that 
all diagonal elements are non-zero. Apart from that we shall give below the explicit form of $F^{-1}$.

To prove the factorization property {\bf A} let us, as above, represent the arbitrary
permutation $\s$ in the form the composition of $k$ elementary transpositions $\s_i$ 
i.e.
$$
\s=\s_{1}...\s_{k}.
$$
The important structural feature of equation (\ref{F}) is that it can be decomposed 
stepwise
into elementary transpositions:
$$
F_{\s(1)...\s(N)}\; R_{1...N}^{\s}=     
$$
$$
=F_{\s_1\s_2...\s_k(1,\ldots,N)}\;
R_{\s_1\s_2...\s_{k-1}(1,\ldots,N)}^{\s_k}\;
R_{\s_1\s_2...\s_{k-2}(1,\ldots,N)}^{\s_{k-1}}
\ldots
R_{1\ldots N}^{\s_1}
$$
$$
=F_{\s_1\s_2...\s_{k-1}(1,\ldots,N)}\;
R_{\s_1\s_2...\s_{k-2}(1,\ldots,N)}^{\s_{k-1}}\;
R_{\s_1\s_2...\s_{k-3}(1,\ldots,N)}^{\s_{k-2}}
\ldots
R_{1\ldots N}^{\s_1}
$$
$$
=\ldots\ldots\ldots=
F_{\s_1(1,\ldots,N)}\;R_{1\ldots N}^{\s_1}\;=\;F_{1...N}
$$
where the composition law (\ref{complaw}) was used.
So we have to prove equation (\ref{F}) for elementary transpositions only.\\
Let $\s_i$ be the elementary transposition $\{i,i+1\}\rightarrow \{i+1,i\}$.
We consider the product $F_{1\ldots i+1\;i\ldots N}R^{\s_i}_{1\ldots N}$. With the help of Eq.'s (\ref{F}) and (\ref{complaw}) we obtain
\beqa
F_{1\ldots i+1\;i\ldots N}R^{\s_i}_{1\ldots N}&=&F_{\s_i(1\ldots i\;i+1\ldots N)}R^{\s_i}_{1\ldots N}\nonumber\\
&=&\sum_{\s\in S_N}\sum^{\quad\quad *(i)}_{\a_{\s_i\s(1)}\ldots\a_{\s_i\s(N)}}
\prod_{j=1}^N P^{\s_i\s(j)}_{\a_{\s_i\s(j)}} R_{\s_i(1,\ldots,N)}^{\s}
R^{\s_i}_{1\ldots N}\nonumber\\
&=&\sum_{\s\in S_N}\sum^{\quad\quad *(i)}_{\a_{\s_i\s(1)}\ldots\a_{\s_i\s(N)}}
\prod_{j=1}^N P^{\s_i\s(j)}_{\a_{\s_i\s(j)}} 
R^{\s_i\s}_{1\ldots N}
\label{FR1}
\eeqa
with the $\sum^{*(i)}$ being defined by the restricting conditions 
\beq
&\a_{\s_i\s(j+1)}\geq\a_{\s_i\s(j)}\quad \mbox{if}\quad\s(j+1)>\s(j)&\nonumber\\
&\a_{\s_i\s(j+1)}>\a_{\s_i\s(j)}\quad \mbox{if}\quad\s(j+1)<\s(j)&\,.
\label{cond(1)}
\eeq
(It may be helpful to keep in mind that the ordering prescription has to be executed according to the shifted labels ${\widetilde{j}}=\s_i(j)$.)
Substituting in (\ref{FR1}) $\tilde{\s}$ for $\s_i\s$ one arrives at
\beqa
F_{1\ldots i+1\;i\ldots N}R^{\s_i}_{1\ldots N}&=&
\sum_{\tilde{\s}\in S_N}\sum^{\quad\quad *}_{\a_{\tilde{\s}(1)}\ldots\a_{\tilde{\s}(N)}}
\prod_{j=1}^N P^{\tilde{\s}(j)}_{\a_{\tilde{\s}(j)}} 
R^{\tilde{\s}}_{1\ldots N}
\label{FR2}
\eeqa
with the defining restrictions of $\sum^{*}$ now of the form
\beq
&\a_{\tilde{\s}(j+1)}\ge\a_{\tilde{\s}(j)}\quad \mbox{if}\quad\s_i\tilde{\s}(j+1)>\s_i\tilde{\s}(j)&\nonumber\\
&\a_{\tilde{\s}(j+1)}>\a_{\tilde{\s}(j)}\quad \mbox{if}\quad\s_i\tilde{\s}(j+1)<\s_i\tilde{\s}(j)&
\label{cond2}
\eeq
which has a slightly different appearance in comparison to (\ref{cond}). Elementary combinatorical considerations lead to the conclusion that the stipulations (\ref{cond}) and (\ref{cond2}) give the same result as long as $\s^{-1}(i)$ and $\s^{-1}(i+1)$ do not happen to be on neighbouring places, that is if not
\beqa
\s^{-1}(i)=\s^{-1}(i+1)\pm 1\,.\label{np}
\eeqa 
If (\ref{np}) holds we have to appeal to the specific form of the $R$-matrix to complete the argument.\\
Comparing the r.h.s. of Eq. (\ref{FR2}) in connection with (\ref{cond2}) to the r.h.s. of Eq. (\ref{F}) in connection with (\ref{cond}) one notes that a discrepancy is certainly excluded if the strict inequality is implied in the step from $\s^{-1}(i)$ to $\s^{-1}(i+1)$ (if $\s^{-1}(i+1)$ is larger than $\s^{-1}(i)$), or from $\s^{-1}(i+1)$ to $\s^{-1}(i)$ if the reversed order is assumed. But for equal group labels at the two neighbouring places in question the representation of the additional transposition of $i$ and $i+1$ in (\ref{F}) as compared to (\ref{FR2}) has no effect, since it supplies a unit factor due to the projectors.\\
It completes the proof of the proposition.\\\\
{\underline{Remark}}: The most general matrix ${\tilde{F}}$ satisfying the 
above conditions {\bf{A}} and {\bf{C}} differs from the special solution of 
the preceding theorem by a non-degenerate, completely symmetric matrix factor \cite{ms},
\beqan
{\tilde{F}}_{1\ldots N}(z_1,\ldots,z_N)&=&X_{1\ldots N}(z_1,\ldots,z_N)F_{1\ldots N}
(z_1,\ldots,z_N),\nonumber\\
X_{1\ldots N}(z_1,\ldots,z_N)&=& X_{\sigma{(1)}\ldots \sigma{(N)}}
(z_{\sigma{(1)}},\ldots,z_{\sigma{(N)}})\;\;\forall \sigma\in S_N\,.
\eeqan
Indeed it is easy to see that ${\tilde{F}}$ satisfies together with $F$ 
the factorization equation (\ref{factor}). Conversely, suppose that both 
$F$ and ${\tilde{F}}$ satisfy (\ref{factor}). It follows that 
\beqan
F^{-1}_{\sigma{(1)}\ldots \sigma{(N)}}F_{1\ldots N}=
{\tilde{F}}^{-1}_{\sigma{(1)}\ldots \sigma{(N)}}{\tilde{F}}_{1\ldots N}
\eeqan
and therefrom
\beqan
F_{1\ldots N}{\tilde{F}}^{-1}_{1\ldots N}=F_{\sigma{(1)}\ldots 
\sigma{(N)}}{\tilde{F}}^{-1}_{\sigma{(1)}\ldots \sigma{(N)}}.
\eeqan
Hence it follows that 
\beqan
X_{1\ldots N}=F_{1\ldots N}{\tilde{F}}^{-1}_{1\ldots N}
\eeqan
is nondegenerate and completely symmetric and transforms $\tilde{F}$ into 
$F$, $X_{1\ldots N}{\tilde{F}}_{1\ldots N}=F_{1\ldots N}$.\\\par\noindent
We need furthemore the inverse operator $F^{-1}$.
To find its expression we have to prove the following 

{\bf Proposition 3.2} The operator $F^*$ defined by the formula
\be
F^*_{1\ldots N} = \sum_{\s\in S_N}\sum^{\quad\quad {**}}_{\a_{\s(1)}\ldots\a_{\s(N)}}
 R_{1\ldots N}^{(t)\;\s}(z_{1},\ldots,z_{N}) \prod_{i=1}^N P^{\s(i)}_{\a_{\s(i)}}
\label{F*}
\ee
with the shorthand notation 
\beqa
R_{1\ldots N}^{(t)\;\s}\equiv R^{\s^{-1}}_{\s(1,\ldots,N)} \label{Rtransp}
\eeqa
and $\sum^{**}$ is taken over all possible 
$\a_{i}$ which satisfy one of two inequalities for each neighbouring pair 
of spaces $i$ and $i+1$:
\beq
&\a_{\s(i+1)}\le\a_{\s(i)}\quad \mbox{if}\quad\s(i+1)<\s(i)&\nonumber\\
&\a_{\s(i+1)}<\a_{\s(i)}\quad \mbox{if}\quad\s(i+1)>\s(i)&
\label{cond1}
\eeq
satisfy the relation
\be
F_{1...N}F^*_{1...N} = \prod_{i<j}\Delta_{ij}
\label{FF*}
\ee
where the diagonal matrix
\be
\label{delta}
{[\Delta_{ij}]}_{\a_i,\a_j}^{\b_i,\b_j}=\d_{\a_i\b_i}\d_{\a_j\b_j}
 \left\{
\begin{array}{ll}
1 \;\;\;\quad \mbox{if}\; \a_i=\a_j,\\
b_{ij} \;\quad \mbox{if}\; \a_i>\a_j,\\
b_{ji} \;\quad \mbox{if}\; \a_j>\a_i
\end{array}
\right.
\ee
acts in the pair of spaces $i$ and $j$.

{\bf Proof} Taking into account the conditions (\ref{cond}) and (\ref{cond1}) in
sums $\sum^*$ and $\sum^{**}$ of the expressions (\ref{F}) and (\ref{F*})
respectively one can write down the expression for 
the product $F_{1...N}F^*_{1...N}$ in the following form:
$$
{F_{1...N}F^*_{1...N}}=
\sum_{\s\in S_N}\sum_{\s'\in S_N}
\sum^{\quad\quad *}_{\a_{\s(1)}\ldots\a_{\s(N)}}
\sum^{\quad\quad {**}}_{\b_{\s'(1)}\ldots\b_{\s'(N)}}
\prod_{i=1}^N P^{\s(i)}_{\a_{\s(i)}} R_{1\ldots N}^{\s}
R^{\s'^{-1}}_{\s'(1)\ldots \s'(N)}\prod_{i=1}^N P^{\s'(i)}_{\b_{\s'(i)}}
$$
\be
=\sum_{\s\in S_N}\sum_{\s'\in S_N}
\sum^{\quad\quad *}_{\a_{\s(1)}\ldots\a_{\s(N)}}
\sum^{\quad\quad {**}}_{\b_{\s'(1)}\ldots\b_{\s'(N)}}
\prod_{i=1}^N P^{\s(i)}_{\a_{\s(i)}} R^{{\s'}^{-1}\s}_{\s'(1,\ldots, N)}
\prod_{i=1}^N P^{\s'(i)}_{\b_{\s'(i)}}
\label{FF*a1}
\ee
\be
=\sum_{\s\in S_N}\sum^{\quad\quad *}_{\a_{\s(1)}\ldots\a_{\s(N)}}
\prod_{i=1}^N P^{\s(i)}_{\a_{\s(i)}} 
R_{\s(N,\ldots,1)}^{\ov\s} 
\prod_{i=1}^N P^{\s(i)}_{\a_{\s(i)}} 
\label{FF*a}
\ee
where the permutation $\ov\s$ reverses the order of the labels:
$$
\ov\s(1,\ldots,N)=(N,\ldots,1).
$$
In the line above (\ref{FF*a1}) we have inserted the definitions of $F$ and $F^*$, Eq's (\ref{F}) and (\ref{F*}) resp. Equality (\ref{FF*a1}) is obtained by applying the compositon rule (\ref{complaw}). To prove equality (\ref{FF*a}) we note first of all that any matrix $R^{\s}$ provides maps s.t. the sets of $sl(n)$ labels of the incoming and outgoing states are connected by a permutation. (This property is easily verified for matrices $R^{\s}$ corresponding to elementary transpositions and it is preserved under the composition of several transpositons.) But the labels $\left\{\a_{\s(i)}\right\}$ represent according to the prescription (\ref{cond}) a non-decreasing series (in ($i$)) of labels whereas the $\left\{\beta_{\s'(i)}\right\}$ - being related to $\left\{\a_{\s(i)}\right\}$ by a permutation - are according to (\ref{cond1}) a non-increasing series. For these two requirements to be fulfilled the equalities
\beqa
\beta_{\s'(N)}=\a_{\s(1)},\;\ldots,\;\beta_{\s'(1)}=\a_{\s(N)}\label{ba}
\eeqa
are a necessity.\\
Let us assume momentarily that all the labels $\beta_{\s'(i)}$ (and hence the $\a_{\s(i)}$) are different from each other. We want to show that Eq. (\ref{ba}) implies the equality 
\beq
\s\,\bar{\s}=\s'\label{ss'}
\eeq 
for the matrix element
\beqan
\left(R^{{\s'}^{-1}\s}_{\s'(1,\ldots, N)}\right)^{\a_{\s(N)}\ldots\a_{\s(1)}}_{\b_{\s'(N)}\ldots\b_{\s'(1)}}
\eeqan
to be non-vanishing. Viewing $R^{{\s'}^{-1}\s}_{\s'(1,\ldots, N)}$ as a product of elementary $R$-matrices one observes that the group label $\beta_{\s'(N)}=\a_{\s(1)}$ can be transported from the lower left corner to the upper right place only if the space labels  $\s'(N)$ and $\s(1)$ are identical. Assume to the contrary that $\s'(N)$ is identical to some other element $\s(x)\neq\s(1)$. It would follow that the group label $\b_{\s'(N)}$ could appear in the upper row only at the place with space label $\s(x)$ or on the l.h.s. of it. (This restriction on the flow of group labels is a straightforward consequence of the form of the elementary $R$-matrix, Eq. (\ref{R}).) We conclude that we have indeed to identify $\s'(N)$ with $\s(1)$ to obtain a non-vanishing matrix element of $R$. The identification of $\s'(N-1)$ with $\s(2)$ etc. follows analogously and therefrom Eq. (\ref{ss'}).\\
A glance on (\ref{cond}) and (\ref{cond1}) affirms that Eq. (\ref{ss'}) remains valid under general circumstances, i.e., if some group lables $\b_{\s'(i)}$ and therefore $\a_{\s(j)}$ occur repeatedly, since the order of the space labels attached to the same group label is uniquely specified by these prescriptions.\\
One deduces from (\ref{FF*a}) that $FF^*$ is a diagonal matrix. A simple calculation leads to the expression for the diagonal elements quoted in Eq. (\ref{delta}). (The product appearing on the r.h.s. of Eq. (\ref{FF*}) is related to $\bar{\s}$  as the latter is a maximal element of $S_N$ and as such is representable as a product of $N(N-1)/2$ elementary transpositions. Each transposition is reflected in one factor of the product in Eq. (\ref{FF*}).)\\
This completes the proof of proposition 3.2.\\
We get from the formula (\ref{FF*}) the expression for $F_{1...N}^{-1}$:
\be
F_{1...N}^{-1}=F^*_{1...N}\prod_{i<j}\Delta^{-1}_{ij}.
\label{Finv}
\ee
For the case of the $sl(2)$ Yangian the formula (\ref{Finv}) corresponds to the result of
proposition 4.6 of \cite{ms}. 

\section{$sl(n)$ generators and the monodromy matrix in the F-basis}

We will first determine the simple root $sl(n)$ generators ${\tilde{E}}_{\a,\a\pm1} = F_{1\ldots N} E_{\a\pm\a +1} F_{1\ldots N}^{-1}$ and the element ${\tilde{T}}_{nn}=F_{1\ldots N} T_{nn} F_{1\ldots N}^{-1}$ of the monodromy matrix. The remaining $sl(n)$ generators can then be obtained from the simple ones through multiple commutators. The examination of the full algebra can be found in Appendix A. One may exploit the $sl(n)$ invariance of the monodromy matrix (with respect to its combined action in the quantum spaces and the auxiliary space, see e.g. \cite{v}) to derive expressions for all elements ${\tilde{T}}_{\a\,\b}$ given ${\tilde{T}}_{nn}$ and the $sl(n)$ generators.\\
One has in particular the relation
\beqa
{\tilde{T}}_{n\a}=\left[{\tilde{E}}_{\a ,n},{\tilde{T}}_{nn}\right]\;.\label{Tna}
\eeqa
The l.h.s. of the latter equation originates from the action of the $sl(n)$ generator in the auxiliary space whereas the r.h.s. evidently reflects the corresponding action in the quantum space.\\
We will content ourselves to derive the explicit form of the ${\tilde{T}}_{n\a}$ using Eq. (\ref{Tna}), since this is all we need to build $sl(n)$ Bethe wave vectors.\\
The simple root generators in the new basis differ from those in the original basis by a diagonal dressing factor. We have\vfill \eject
{\bf Proposition 4.1} 
\be
\tilde E_{\g,\g\pm 1}=\sum_{i=1}^N E^{(i)}_{\g,\g\pm 1}\otimes_{j\ne i}
{G^{\pm\g}(i,j)}_{[j]}
\label{Epm}
\ee
where 
$$
{G^{\g}(i,j)}_{k,l}=\d_{kl}
 \left\{
\begin{array}{ll}
b^{-1}_{ij}\; \mbox{if}\; k=\g,\\
1\;\quad\; \mbox{otherwise}
\end{array}
\right.
$$
\be
\label{Gpm}
{G^{-\g}(i,j)}_{k,l}=\d_{kl}
 \left\{
\begin{array}{ll}
b^{-1}_{ji}\; \mbox{if}\; k=\g+1\\
1\; \quad\;\mbox{otherwise}\;.
\end{array}
\right.
\ee

{\bf Proof}  
Eq's (\ref{Epm}) and (\ref{Gpm}) specialized to the rational $sl(2)$ case have been presented in propositions 5.1 and 5.2 of ref. \cite{ms}. The proof of these equations for the $sl(n)$ model with arbitrary $n$ can be reduced to that of the $sl(2)$ model. One has to note for this purpose that one obtains due to the $sl(n)$ invariance of the elementary R-matrices the vanishing result
\beqa
\left[R^{\s}_{1\ldots N},E_{\a,\a\pm 1}\right]=0
\label{Rinvar}
\eeqa
for any permutation $\s\in S_N$.\\
This allows us to write (cf. Eq. (\ref{FF*a1}))
\beqa
{\tilde{E}}_{\g,\g\pm 1}=\sum_{\s\in S_N}\sum_{\s'\in S_N}
\sum^{\quad\quad *}_{\a_{\s(1)}\ldots\a_{\s(N)}}
\sum^{\quad\quad {**}}_{\b_{\s'(1)}\ldots\b_{\s'(N)}}
\prod_{i=1}^N P^{\s(i)}_{\a_{\s(i)}} E_{\g,\g\pm 1} R^{{\s'}^{-1}\s}_{\s'(1,\ldots, N)}
\prod_{i=1}^N P^{\s'(i)}_{\b_{\s'(i)}}\prod_{i<j}\Delta_{ij}\;.
\label{E1}
\eeqa
The collapse of the double sum $\sum_{\s,\s'}$ into a single sum proceeds here along the same pattern as above (in the transition from Eq. (\ref{FF*a1}) to Eq. (\ref{FF*a})). One further has to note that group indices $\g$ and $(\g +1)$ ($(\g -1)$ resp.) only occur in neighbouring positions what concerns ingoing and outgoing matrix indices because of the monotonicity prescription incorporated into the sums $\sum^*$ and $\sum^{**}$ resp. The rearrangement of the neighbouring labels $\g$ and $(\g +1)$ ($(\g -1)$ resp.) goes on according to $sl(2)$ rules and produces the result quoted in Eq. (\ref{Gpm}) and in ref. \cite{ms}. Rearrangements involving group indices different from $\g$ and $(\g +1)$ ($(\g -1)$ resp.) are not affected by the presence of the generator $E_{\g,\g\pm 1}$, since for those rearrangements the difference of $\g$ and $(\g +1)$ ($(\g -1)$ resp.) is immaterial.\par\noindent
{\bf{Proposition 4.2}}
\be
\tilde T_{nn}(\l) = \otimes_{i=1}^N \mbox{diag}\{b(\l-z_i),\ldots,b(\l-z_i),1\}\,.
\label{Tnn}
\ee
{\bf{Proof}} Let us consider the action of the matrix $F$ on $T_{nn}$
\beqa
F_{1\ldots N}T_{nn}&=&\sum_{\s\in S_N}\sum^{\quad\quad *}_{\a_{\s(1)}\ldots\a_{\s(N)}}
\prod_{i=1}^N P^{\s(i)}_{\a_{\s(i)}} R_{1\ldots N}^{\s}P_n^0 T_{0,1\ldots N} P_n^0\nonumber\\
&=&\sum_{\s\in S_N}\sum^{\quad\quad *}_{\a_{\s(1)}\ldots\a_{\s(N)}}
\prod_{i=1}^N P^{\s(i)}_{\a_{\s(i)}}P_n^0 T_{0,\s(1)\ldots \s(N)} P_n^0
 R_{1\ldots N}^{\s}\,.\label{Tnn1}
\eeqa
The specialization to the entry $(n,n)$ of the auxiliary space is here achieved by the projectors $P_n^0$. For the second equality in (\ref{Tnn1}) we have used relation (\ref{lcrperm}) and the obvious fact that $P_n^0$ commutes with $R^{\s}_{1\ldots N}$. To simplify the following argument we distinguish in the sum $\sum^*$ cases of various multiplicities of the occurence of the group index $n$: 
\beqa
F_{1\ldots N}T_{nn}&=&\sum_{\s\in S_N}\sum_{k=0}^N\sum^{\quad\quad *'}_{\a_{\s(1)}\ldots\a_{\s(N)}}
\prod_{j=N-k+1}^N\delta_{\a_{\s(j)},n}\prod_{j=1}^{N-k} P^{\s(j)}_{\a_{\s(j)}}P_n^0 T_{0,\s(1)\ldots \s(N)} P_n^0
 R_{1\ldots N}^{\s}\,.\label{Tnn2}
\eeqa
Let us consider the prefactor of $R^{\s}_{1\ldots N}$ on the r.h.s. of Eq. (\ref{Tnn2}) more closely. Using specific features of the $R$-matrices we can rewrite it as follows:
\beqa
&&\prod_{j=1}^{N-k} P^{\s(j)}_{\a_{\s(j)}}\prod_{j=N-k+1}^N P_n^{\s(j)} P_n^0\,T_{0,\s(1)\ldots \s(N)}\,P_n^0\nonumber\\
&=&\prod_{j=1}^{N-k} P^{\s(j)}_{\a_{\s(j)}}\left(R_{0,\s(N)}\right)_{nn}^{nn} \left(R_{0,\s(N-1)}\right)_{nn}^{nn}\ldots\left(R_{0,\s(N-k+1)}\right)_{nn}^{nn}P_n^0\,T_{0,\s(1)\ldots \s(N-k)}\,P_n^0\prod_{j=N-k+1}^N P_n^{\s(j)}\nonumber\\
&=&\prod_{j=1}^{N-k} P^{\s(j)}_{\a_{\s(j)}}P_n^0\,T_{0,\s(1)\ldots \s(N-k)}\,P_n^0 \prod_{j=N-k+1}^N P_n^{\s(j)}\nonumber\\
&=&\prod_{i=1}^{N-k}\left(R_{0i}\right)_{n,\a_{\s(i)}}^{n,\a_{\s(i)}}\prod_{j=1}^{N-k} P^{\s(j)}_{\a_{\s(j)}}\prod_{j=N-k+1}^N P_n^{\s(j)}\label{prp}
\eeqa
Inserting the r.h.s. of (\ref{prp}) into Eq. (\ref{Tnn2}) one sees that the product $\prod_i\left(R_{0i}\right)_{n,\a_{\s(i)}}^{n,\a_{\s(i)}}$ provides the desired diagonal dressing factor of $T_{nn}$ and the product of projectors applied to $R^{\s}$ restores $F_{1\ldots N}$.\\
This completes the proof of proposition 4.2.\par\noindent
Given the simple root generators $\tilde E_{\a,\a\pm 1}$ it is a straightforward task to evaluate the generators corresponding to non-simple roots.\\
One finds in particular 
\beqa
\tilde E_{n-\a,n}=\sum_{k=1}^{\a}\sum_{i_1\ne\ldots\ne i_k}
\prod_{\g=1}^{k-1}
{{\eta}
\over
{z_{i_{\g}}-z_{i_{\g+1}}}
}\sum_{\a=\b_0>\b_1\ldots >\b_k=0} \otimes_{l=1}^{k} E^{(i_{l})}_{n-\b_{l-1},n-\b_{l}}
\otimes_{j\ne i_1\ldots i_k} 
\Gamma^{(j)}_{j;{\underbrace{i_1..i_1}_{\b_0-\b_1}}\;
{\underbrace{i_2..i_2}_{\b_1-\b_2}}\ldots {\underbrace{i_k..i_k}_{\b_{k-1}-\b_k}}}
\label{Ean}
\eeqa
where $\Gamma_{j;i_1,\ldots,i_{\a}}=\mbox{diag}\{1,...,1,b^{-1}_{i_1j},...,
b^{-1}_{i_{\a}j},1\}$.\\\\
Exploiting the last equation and Eq. (\ref{Tna}) one finally arrives at
\beqa
\tilde T_{n\;n-\a}=\sum_{k=1}^{\a}\sum_{i_1\ne\ldots\ne i_k}
c(\l-z_{i_k})\prod_{\g=1}^{k-1}
{{\eta}
\over
{z_{i_{\g}}-z_{i_{\g+1}}}
} b(\l-z_{i_{\g}})\;\,\times\nonumber\\
\
\sum_{\a=\b_0>\b_1\ldots >\b_k=0} \otimes_{l=1}^{k} E^{(i_{l})}_{n-\b_{l-1},n-\b_{l}}
\otimes_{j\ne i_1\ldots i_k} 
\Delta^{(j)}_{j;{\underbrace{i_1..i_1}_{\b_0-\b_1}}\;
{\underbrace{i_2..i_2}_{\b_1-\b_2}}\ldots {\underbrace{i_k..i_k}_{\b_{k-1}-\b_k}}}
\label{TTna}
\eeqa
where $\Delta^{(j)}_{j;i_1,\ldots,i_{\a}}=\mbox{diag}\{b(\l-z_j),...,b(\l-z_j),
b(\l-z_j)\,b^{-1}_{i_1j},...,
b(\l-z_j)\,b^{-1}_{i_{\a}j},1\}$ is a diagonal dressing matrix acting in $j$-th space.

\section{Bethe wave vectors}

What concerns the description of the hierarchical Bethe ansatz we will be rather sketchy, referring for more details to \cite{kulresh} and \cite{v}.\\
The operators $T_{n\alpha}(\la)\;(1\leq\alpha<n-1)$ serve in the $sl(n)$ problem as quasiparticle creation operators and the corresponding operators $T_{\alpha n}(\la)$ have the role of annihilation operators.\\
The $T_{n\alpha}(\la)$ satisfy the Faddeev--Zamolodchikov algebra 
$$
[T_{n\a}(\l_1),T_{n\a}(\l_2)]=0
$$
\be
T_{n\a}(\l_1)T_{n\b}(\l_2)={{1}\over{b(\l_2-\l_1)}} T_{n\b}(\l_2)T_{n\a}(\l_1)
-{{c(\l_2-\l_1)}\over{b(\l_2-\l_1)}} T_{n\b}(\l_1)T_{n\a}(\l_2)
\label{algebra}
\ee
where in the last relation $\a\ne\b$.\\
An ansatz for a Bethe vector $\Psi_n$ is given in terms of a linear superposition of products of operators $T_{n\a}$ acting on a reference state $\Omega ^{(n)}_N$:
\be
\Psi_n (N;\l_1,\ldots,\l_p)=\sum_{\a_1,\ldots,\a_{p}}
\Phi_{\a_1,\ldots,\a_{p}}
T_{n\a_1}(\l_1)\ldots T_{n\a_{p}}(\l_{k})\,\Omega^{(n)}_N
\label{Psi_n}
\ee
where the  reference state $\Omega^{(n)}_N$ is constituted as a $N$-fold tensor product of lowest weight states $v_n^{(i)}$ in ${\mathbb{C}}_n^{(i)}$
\beqan
\Omega_N=\otimes_{i=1}^N v_n^{(i)}
\eeqan
and the $\Phi_{\a_1,\ldots,\a_{p}}$ denote some c-number coefficients.\\
It is important to note that the reference state is invariant under the $F$-transformation:
\beqan
F\,\Omega^{(n)}_N=\Omega^{(n)}_N
\eeqan
since it is immediate from the definition (\ref{F}) of $F$ that from the sum over the permutation group only the term with the unit element inscribed gives a non-vanishing result when applied to $\Omega^{(n)}_N$.\par\noindent
It can be shown, \cite{kulresh}, \cite{v}, that $\Psi_n$ is eigenvector of the transfer matrix $t(\l)=\sum_{i}T_{ii}(\l)$ if\\
$i)$ the parameters $\l_1,\ldots,\l_p$ satisfy a certain system of rational equations, the famous Bethe ansatz equations\\
and if \\
$ii)$ the c-number coefficients are chosen s.t. they constitute the components of a rational $sl(n-1)$ transfer matrix.\\
One establishes therewith a recursive procedure leading finally to a $sl(2)$ eigenvalue problem. We will keep the spectral parameters arising in the various stages of the procedure in general position  instead of specializing them to solutions of the Bethe ansatz equations. We keep in other words the Bethe vector ``off-shell'' \cite{bab}. Our goal in this paper is to figure out the functional form of the Bethe wave vectors.\par\noindent
To start with we recall the form of the $sl(2)$ wave vectors in the basis provided by Maillet and Sanchez de Santos \cite{ms}. The creation operators with respect to the lowest weight reference state  (in the special basis) are of the form
\be
{\tilde T}_{21}(\l)=\sum_{i=1}^N c(\l-z_i)\s_+^{(i)}
\otimes_{j\ne i}
{\left(\matrix{
b(\l-z_j)b_{ij}^{-1}&0\cr
0\quad\quad&1\cr
}\right)}_{[j]}
\label{T21}
\ee
The ensuing Bethe wave vectors are given by
\beqa
\Psi_2(N;\l_1,\ldots,\l_p)&=&\tilde{T}_{21}(\l_1)\ldots \tilde{T}_{21}(\l_p)\,\Omega^{(2)}_N\nonumber\\
&=&\sum_{i_1\ne\ldots\ne i_{p}} B_{p}^{(2)}(\l_1,\ldots,\l_{p}|z_{i_1},\ldots,z_{i_{p}})
\s_+^{(i_1)}\ldots\s_+^{(i_{p})}\,\Omega^{(2)}_N\;.
\label{Psi_2}
\eeqa
The c-number coefficients $B^{(2)}(\{\l_i\}|\{z_i\})$ of the last equation can easily be worked out - taking into account the ``diagonal dressing'' factors of the spin raising operators $\s_{+}^{i}$ in (\ref{T21}) - to be of the form 
\beqa
B^{(2)}_p(\l_1,\ldots,\l_p|z_{1},\ldots,z_{p})&=&
\sum_{\s\in S_p}\prod_{m=1}^p c(\l_m-z_{\s(m)})
\prod_{l=m+1}^{p}{{b(\l_{m}-z_{\s(l)})}\over{b(z_{\s(m)}-z_{\s(l)})}}.
\label{B_p1}
\eeqa
A concise alternative representation of the coefficients $B_{p}^{(2)}$ has been derived in \cite{maillet}:
\be
B_{p}^{(2)}(\l_1,\ldots,\l_p|z_{i_1},\ldots,z_{i_p})=
{{\prod_{i,j}(\l_i-z_j)}\over{\prod_{i>j}(\l_i-\l_j)(z_j-z_i)}}
\det_{<i,j>}({1\over{\l_i-z_j}}-{1\over{\l_i-z_j+\eta}}).
\label{B_p}
\ee
The vectors $\Psi^{(2)}_p(N;\l_1,\ldots,\l_p)$ are invariant under arbitrary exchanges of the variables $\l_1,\ldots,\l_p$ since operators $\tilde{T}_{21}$ with different values of the attached spectral parameters do commute with each other.\\
It has been shown in \cite{v}, \cite{takht} that this symmetric appearance of the spectral parameters in the wave vectors $\Psi^{(n)},\;n>2$ persists - despite of the Faddeev-Zamolodchikov relations (\ref{algebra}) - under the assumption that the coefficients $\Phi_{\a_1,\ldots,\a_{p}}$ in (\ref{Psi_n}) are components of a $sl(n-1)$ Bethe wave vector. Our argumentation below will heavily rely on this exchange symmetry.\par\noindent
We discuss now the $sl(3)$ model. The generalization to $sl(n),\;n>3$ will  afterwards be rather obvious. Eq. (\ref{TTna}) specialized to the case of $sl(3)$ renders the creation operators in the $F$-basis as
\beqa
{\tilde T}_{32}\;&=&\;\sum_{i=1}^N c(\l-z_i) E_{23}^{(i)}\otimes_{j\neq i}
\mbox{diag}\{
b(\l-z_j),b(\l-z_j)b^{-1}_{ij},1\}_{[j]}\label{T_32}\\
{\tilde T}_{31}\;&=&
\;\sum_{i=1}^N c(\l-z_i) E_{13}^{(i)}\otimes_{j\neq i}
\mbox{diag}\{
b(\l-z_j)b^{-1}_{ij},b(\l-z_j)b^{-1}_{ij},1\}_{[j]}\;\,+\nonumber\\
\sum_{i\neq j} c(\l-z_i)&&\hspace{-2em}b(\l-z_j)\,{{\eta}\over{z_i-z_j}}E_{23}^{(i)}\otimes
E_{12}^{(j)}
\otimes_{k\neq i,j}
\mbox{diag}\{
b(\l-z_k)b^{-1}_{jk},b(\l-z_k)b^{-1}_{ik},1\}_{[k]}.\label{T_31}
\eeqa
The strategy employed in determining the form of the Bethe wave vector (\ref{Psi_n}) will be as follows:\\
-- We select a particular order in which the operators $T_{n\a}$ act on the reference state s.t. the eventual explicit evaluation becomes as simple as possible. (This particular order can always be achieved by the use of the Faddeev--Zamolodchikov relations (\ref{algebra}).)\\
-- The c-number coefficient $\Phi^{(2)}$ has to be taken in the original basis  and not in the $F$ basis, but fortunately a particular coefficient in the sum (\ref{Psi_n}) (specialized to $sl(3)$) is invariant under the similarity transformation induced by the $F$-matrices. This enables one to compute the explicit form of this special coefficient $\Phi^{(2)}$ using the result (\ref{B_p1}) and relate it to the order of operators alluded to in the preceding point by an appropriate factor $\prod_{ij}b^{-1}(\l_i-\mu_j)$.\\
-- One uses the permutation symmetry to determine all other terms in the sum.\\\\
Following this line of thought we arrive at the following\\
\vspace{0.2cm}
{\bf Proposition 5.1}
\beqa
&&{\tilde\Psi}_3(N,\l_1,\ldots,\l_{p_0};\l_{p_0+1},\ldots,\l_{p_0+p_1})= \nonumber\\
&&\sum_{\s \in S_{p_0}} B^{(2)}_{p_1}(\l_{p_0+1},\ldots,\l_{p_0 +p_1}|
\l_{\s(1)},\ldots,\l_{\s(p_1)})\prod_{k=1}^{p_1}\prod_{l=p_1+1}^{p_0}
b(\l_{\s(k)}-\l_{\s(l)})^{-1}\nonumber\\
&&{\tilde T}_{32}(\l_{\s(p_1 +1)})\ldots {\tilde T}_{32}(\l_{\s(p_0)})
{\tilde T}_{31}(\l_{\s(1)})\ldots {\tilde T}_{31}(\l_{\s(p_1)})\,\Omega^{(3)}_N
\label{Psi_3a}
\eeqa

\vspace{0.2cm}
{\bf Proof} $\quad$
The proof of this formula procceeds as mentioned above:\\
We have specialized the form of the ansatz in Eq. (\ref{Psi_3a}) as compared to Eq. (\ref{Psi_n}) so that operators $\tilde{T}_{32}$ are placed to the left of all operators $\tilde{T}_{31}$. The latter order can be achieved by moving the operators $\tilde{T}_{32}$ in the general ansatz (\ref{Psi_n}) to the wanted position with the help of the Faddeev--Zamolodchikov  relations (\ref{algebra}). Let us consider in particular the vector contributing in  (\ref{Psi_n}) of the form 
\beqa
{\tilde T}_{31}(\l_1)\ldots {\tilde T}_{31}(\l_{p_1})
{\tilde T}_{32}(\l_{p_1+1})\ldots {\tilde T}_{32}(\l_{p_0})\,\Omega_N^{(3)}\label{12}
\eeqa
and let us relate it to the vector contributing in  (\ref{Psi_3a}) of the form 
\beqa
{\tilde T}_{32}(\l_{p_1+1})\ldots {\tilde T}_{32}(\l_{p_0})
{\tilde T}_{31}(\l_1)\ldots {\tilde T}_{31}(\l_{p_1})\,\Omega_N^{(3)}.\label{21}
\eeqa
A diligent appreciation of the Faddeev--Zamolodchikov relations leads to the conclusion that (\ref{21}) has its unique origin in (\ref{12}) and that moreover only the first term on the r.h.s. of (\ref{algebra}) supplies  contributions in the transition from (\ref{12}) to (\ref{21}).  It follows that the  transition from (\ref{12}) to (\ref{21}) is accompanied by an additional factor
\beqa
\prod_{x=1}^{p_1}\prod_{y=p_1+1}^{p_0}{{1}\over{b(\l_{x}-\l_{y})}}\label{factorb}
\eeqa
The c-number coefficients  $\Phi_{\a_1\ldots \a_p}$ in (\ref{Psi_n}) (when specialized to the case $n=3$) refer to a $sl(2)$ Bethe wave vector in the familiar basis commonly used for the algebraic Bethe ansatz - not the one of Maillet and Sanchez de Santos.\\
But we want to argue that the special coefficient $\Phi^{(2)}_{1\ldots 12\ldots 2}$ (the factor which accompanies the vector (\ref{21})) is in fact the same in both frames. One has to note first that the similarity transformation by the $F$-matrices (specialized to the case of $sl(3)$) respects the $sl(2)$ structure. This means among other things that components only with the same number of labels $1$ and $2$ are related to each other through the similarity transformation. One has secondly to observe that in the transformation of $\Phi^{(2)}_{1\ldots 12\ldots 2}$ no other components with a different order of labels can appear due to the lower triangularity of $F$. (The matrix $F$ would otherwise not be lower triangular).\\
One finds thirdly through a direct examination of the definition of F that its diagonal elements relating the coefficients $\Phi^{(2)}_{1\ldots 12\ldots 2}$ in the two frames to each other is equal to unity. Therefore we know the coefficient $\Phi^{(2)}_{1\ldots 12\ldots 2}$ to be of the  Maillet -- Sanchez de Santos form.\\
Invoking the above mentioned exchange symmetry one completes the proof.\par\noindent
The expression (\ref{Psi_3a}) for $\tilde{\Psi}_3$ can be worked out further by inserting the definitions (\ref{T_31}) and  (\ref{T_32}) of $\tilde{T}_{31}$ and $\tilde{T}_{32}$ resp. to yield
$${\tilde\Psi}_3(N,\l_1,\ldots,\l_{p_0};\l_{p_0+1},\ldots,\l_{p_0+p_1})
=
\sum_{i_1\ne\ldots\ne i_{p_0}} B_{p_0,p_1}^{(3)}(\l_1,\ldots,\l_{p_0};\l_{p_0+1},\ldots,
\l_{p_0+p_1}|z_{i_1},\ldots,z_{i_{p_0}})
$$
\be
E_{23}^{(i_{p_1+1})}\ldots E_{23}^{(i_{p_0})} E_{13}^{(i_{1})}\ldots 
E_{13}^{(i_{p_1})} \Omega^{(3)}_N
\label{Psi_3b}
\ee
The order of operators adopted in Eq. (\ref{Psi_3a}) yields the bonus that the second term on the r.h.s. (the twofold sums) of (\ref{T_31}) do not appear in (\ref{Psi_3b}), since those drop out if applied to the reference state $\Omega_N$.\\
The sets of operators $\tilde{T}_{31}$ and $\tilde{T}_{32}$ generate through their respective diagonal dressing the structure of two $sl(2)$ wave vectors together with a factor which accounts for the way the operators  $\tilde{T}_{32}$ respond to operators $\tilde{T}_{31}$ on their right hand side (cf. Eq. (\ref{12})).\\
This completes our goal to reduce the $sl(3)$ Bethe wave vectors to $sl(2)$ structures: 
$$
B_{p_0,p_1}^{(3)}(\l_1,\ldots,\l_{p_0};\l_{p_0+1},\ldots,
\l_{p_0+p_1}|z_{i_{1}},\ldots,z_{i_{p_0}})=
\qquad\qquad
$$
$$
\sum_{\s \in S_{p_0}}
\prod_{k=1}^{p_1}\prod_{l=p_1+1}^{p_0}
{{b(\l_{\s(l)}-z_{i_k})}\over{b(\l_{\s(k)}-\l_{\s(l)})}}
B_{p_0-p_1}^{(2)}(\l_{\s(p_1+1)},\ldots,\l_{\s(p_0)}|z_{i_{p_1+1}},\ldots,z_{i_{p_0}})
$$
\be
B_{p_1}^{(2)}(\l_{p_0+1},\ldots,\l_{p_0+p_1}|\l_{\s(1)},\ldots,\l_{\s(p_1)}) 
B_{p_1}^{(2)}(\l_{\s(1)},\ldots,\l_{\s(p_1)}|z_{i_{1}},\ldots,z_{i_{p_1}})
\label{B1}
\ee
\par\noindent
All ingredients of our argumentation for the case of $sl(3)$ can be straightforwardly generalized to $sl(n);\;\;n>3$.\\ 
We collect all operators $\tilde{T}_{n\,n-\alpha}$ to the left of operators $\tilde{T}_{n\,n-\beta}$ if $\alpha<\beta$. Once again only the first term in the expression (\ref{TTna}) of the respective operators $\tilde{T}_{n\,n-i}$ contributes in this special ordering.\\
The wave function $\tilde{\Psi}_n$ is then  expressed in analogy to Eq. (\ref{Psi_3b}) by:
$$
\tilde{\Psi}_n(N,p_0,p_1,\ldots,p_{n-2})=
$$
\be
\sum_{i_1\ne\ldots\ne i_{p_0}} B^{(n)}_{p_0,p_1,\ldots,p_{n-2}}
(\l_1,\ldots,\l_{p_0+...p_{n-2}}|z_{i_1},\ldots,z_{i_{p_0}})
\prod_{\a=1}^{n-1}\prod_{j=p_{\a}+1}^{p_{\a-1}}
E_{n-\a\;n}^{(i_j)} \Omega^{(n)}_N
\label{Psi_na}
\ee
with the following recursion relation for the function $B^{(n)}$:
$$
B^{(n)}_{p_0\,p_1\,\ldots\,p_{n-2}}(\l_1,\ldots,\l_{p_0+p_1+\ldots p_{n-2}}|z_{1},\ldots,z_{{p_0}})
$$
$$
=\sum_{\s\in S_{p_0}}\prod_{\a=1}^{n-2}\quad\prod_{k_{\a}=p_{\a+1}+1}^{p_{\a}}
\quad\prod_{l_{\a}=p_{\a}+1}^{p_0}
{
{b(\l_{\s(l_{\a})}-z_{{k_{\a}}}) }\over
{b(\l_{\s(k_{\a})}-\l_{\s(l_{\a})})}
}
$$
$$
\prod_{\g=0}^{n-2} 
B^{(2)}_{p_{\g}-p_{\g+1}}(\l_{\s(p_{\g+1}+1)}\ldots\l_{\s(p_{\g})}|
z_{{p_{\g+1}+1}}\ldots z_{{p_{\g}}})
$$
\be
B^{(n-1)}_{p_1\ldots p_{n-2}}(\l_{p_0+1}\ldots \l_{p_0+p_1+\ldots +p_{n-2}}|
\l_{\s(1)}\ldots\l_{\s(p_1)})
\label{B_n}
\ee
The resolution of the recursion gives
\beqa
&& B^{(n)}_{p_0\,p_1\,\ldots\,p_{n-2}}(\l_1,\ldots,\l_{p_0+p_1+\ldots +p_{n-2}}|z_{1},\ldots,z_{{p_0}})=\nonumber\\
&&\sum_{\s_0\in S_{p_0}}\sum_{\s_1\in S_{p_1}}\ldots\sum_{\s_{n-3}\in S_{p_{n-3}}}\prod_{i=0}^{n-2}\;\prod_{\a_i=i+1}^{n-2}\;\prod_{k_{\a_i}=p_{\a_i+1}+1}^{p_{\a_i}}\;\prod_{l_{\a_i}=p_{\a_i}+1}^{p_i}\;
{{b(\l_{q_{i-1}+{\s_i(l_{\a_i})}}-\l_{\s_{i-1}(k_{\a_i})})}\over{b(\l_{q_{i-1}+{\s_i(k_{\a_i})}}-\l_{q_{i-1}+{\s_{i}(l_{\a_i})}})}}\nonumber\\
&&\prod_{\gamma_i=i}^{n-2}\,B^{(2)}_{p_{\gamma_i}-p_{\gamma_i+1}}\left(\l_{q_{i-1}+{\s_i(p_{\gamma_{i+1}+1})}}\ldots\l_{q_{i-1}+{\s_i(p_{\gamma_i})}}|\l_{\s_{i-1}(p_{\gamma_{i+1}+1})}\ldots\l_{\s_{i-1}(p_{\gamma_i})}\right)\;\times\nonumber\\
&& B^{(2)}_{p_{n-2}}\left(\l_{q_{n-3}+1}\ldots\l_{q_{n-3}+p_{n-2}}|\l_{q_{n-4}+\s_{n-3}(1)}\ldots\l_{q_{n-4}+\s_{n-3}(p_{n-2})}\right)\label{B_n1}
\eeqa
where by definition 
$$
q_{i}=\sum_{j=0}^{i}p_{j};\;\;q_{-1}=0
$$
and 
$$
\l_{\s_{-1}(k)}=z_k\,.
$$
Eq's (\ref{Psi_na}) and (\ref{B_n1}) supply the explicit representation of the $sl(n)$ wave vectors in terms of $sl(2)$ vectors, that is, the resolution of the Bethe hierarchy.

\section{ Conclusions}

The form of  the factorizing  $F$-matrix presented in section 3  is of an 
intriguing simplicity. We suspect that a representation theoretical aspect 
is lurking behind it which escapes our present knowledge. It should be noted 
that we arrived  at this ansatz by guesswork immediately for the full $F$-matrix 
instead of taking the detour via partial
$F$-matrices, as proposed in \cite{ms}. It seems rather likely that we would have 
missed the simplicity of the ansatz  if we had chosen the approach  via partial 
$F$-matrices.\\
Our original hope was to find  a structure for the Bethe wave vectors which is 
as suggestive as the one displayed for the case of Gaudin magnets in \cite{flume}. 
This goal has not yet been achieved completely  since we are not in possession of 
an entirely satisfactory representation of $sl(2)$ wave vectors, which are the 
building blocks for the 
final formula (\ref{B_n1}) of section 5. The representations (\ref{B_p1}), (\ref{B_p}) both have the drawback that they do display the singularity structure of the wave vectors in a redundant manner. (The matter is further discussed in Appendix B.) We nevertheless  nourish the hope that 
our findings will be of some help to bring effective large $n$ calculations of 
thermodynamical quantities into the range of the algebraic Bethe ansatz method.\\
\vspace{1.5cm}\\
{\bf{Acknowledgement:}} We are indebted to J.--M. Maillet for a seminar and ensuing fruitful discussions, which initiated the present work. H.B. thanks the Alexander von Humboldt Foundation for support. R.F. was supported by the TMR network contract FMRX-CT96-0012 of the European Commission. 

\begin{appendix}
\setcounter{secnumdepth}{-1}
\section{Appendix A}
In this appendix we verify the $sl$(n) algebra relations taking the formulas for the generators $\tilde{E}_{\a,\a\pm 1}$ of section 4 as a starting point.\\\\
We use the following defining relations for a semisimple Lie algebra \cite{hum}:\\
Fix a root system with a basis $\{\alpha_1,\ldots,\alpha_l\}$. Let L 
be the Lie algebra generated by $3\;l$ elements 
$\{{E}_{+\alpha_i},{E}_{-\alpha_i},H_i;\;1\leq i\leq l\}$. 
$L$ is uniquely determined by the relations
\begin{itemize}
\item{S1}\hspace{3em}
$\left[E_{+\alpha_{i}},E_{-\alpha_{j}}\right]=\delta_{ij}H_{\alpha_{i}}$
\item{S2}\hspace{3em}
$\left[H_{\alpha_{i}},E_{\pm\alpha_{j}}\right]=\pm A_{ji} E_{\pm\alpha_{j}}$
\item{S3}\hspace{3em}
$\left[H_{\alpha_{i}},H_{\alpha_{j}}\right]=0$
\item{S4}$\;\,$\footnote{$(ad_x)$ is a shorthand notation for 
${{\underbrace{ad_x\circ ad_x\circ\ldots\circ ad_x}}_{\rm{n times}}}$, 
such that e.g. $(ad_x)^2(y)=[x,[x,y]]$}\hspace{3em} 
$\left(ad_{E_{\pm}^{\alpha_{i}}}\right)^{1-A_{ji}}\left(E_{\pm}^{\alpha_{j}}\right)=0
\;\;\;i=1,\ldots,l;\;\;i\neq j$
\end{itemize}
with $A_{ij}=2{{(\alpha_i,\alpha_j)}\over{(\alpha_j,\alpha_j)}}$ denoting the Cartan matrix. \\\\
We recall from section 4 the expressions for the generators of the algebra $sl(n)$ corresponding to simple roots:
\beq
{\tilde{E}}_{+\alpha}&=&
\sum_{i=1}^N E^{(i)}_{+\alpha}{\otimes_{j\neq i}}
\left(\id_N+{{\eta}\over{z_i-z_j}}e_{\alpha\;\alpha}
      \right)_{[j]}\equiv
\sum_{i=1}^N E^{(i)}_{+\alpha}{\otimes_{j\neq i}}
{\tr}^{\alpha}_{(i,j)}
\nonumber\\
{\tilde{E}}_{-\alpha}&=&
\sum_{i=1}^N E^{(i)}_{-\alpha}{\otimes_{j\neq i}}
\left(\id_N+{{\eta}\over{z_j-z_i}}e_{\alpha+1\;\alpha+1}
      \right)_{[j]}\equiv\sum_{i=1}^N E^{(i)}_{-\alpha}{\otimes_{j\neq i}}
{\tilde{\tr}}^{\alpha}_{(j,i)}
\eeq
where $\left(e_{ij}\right)_{kl}=\delta_{ik}\delta_{jl}$ are the elementary 
matrices and $\left(E^{(k)}_{+\alpha}\right)_{ij}=\delta_{\alpha\;i}
\delta_{\alpha+1\;j},\;\;\left(E^{(k)}_{-\alpha}\right)_{ij}=\delta_{\alpha+1\;i}
\delta_{\alpha\;j}$ are the simple roots of $sl(n)$ acting in the k-th space.
\par\noindent
Using their definitions one has \footnote{$\sum^{\prime}_{i,j}$ means $\sum_{i,j\;i\neq j}$}.
\beq
&&\left[{\tilde{E}}_{+\alpha},{\tilde{E}}_{-\beta}\right]=\nonumber\\
&=&\sum_{i}\left[E^{(i)}_{+\alpha},E^{(i)}_{-\beta}\right]\otimes_{j\neq i}
{\tr}^{\alpha}_{(i,j)}{\tilde{{\tr}}}^{\beta}_{(j,i)}\nonumber\\
&+&\sum_{i,j} {}^{\prime}\left(E_{\alpha}^{(i)}{\tilde{{\tr}}}^{-\beta}_{(i,j)}
\otimes{\tr}^{\alpha}_{(i,j)}E_{-\beta}^{(j)}-{\tilde{{\tr}}}^{-\beta}_{(i,j)}
E_{\alpha}^{(i)}\otimes E_{-\beta}^{(j)}{\tr}^{\alpha}_{(i,j)}\right)
\otimes_{k\neq i,j} {\tr}^{\alpha}_{(i,k)}{\tilde{{\tr}}}^{-\beta}_{(j,k)}=
\nonumber\\
&=&\sum_{i}\delta_{\alpha\;\beta}H^{(i)}_{\alpha}\otimes_{j\neq i}
{\tr}^{\alpha}_{(i,j)}{\tilde{{\tr}}}^{\alpha}_{(j,i)}\nonumber\\
\eeq
where we exploited the fact that the second sum vanishes term by 
term identically for all $\{\alpha,-\beta\}$.\\
The dressing can be written as ${\tr}^{\alpha}_{(i,j)}
{\tilde{{\tr}}}^{\alpha}_{(j,i)}
=\id_{[j]}+{{\eta}\over{z_i-z_j}}H^{(j)}_{\alpha}$, because 
$\left(H_{\alpha}\right)_{ij}=\delta_{\alpha\;i}\delta_{\alpha\;j}-
\delta_{\alpha+1\;i}\delta_{\alpha+1\;j}$.\\
Against first appearance the Cartan operators $H_{\a}$ remain without dressing. 
For this purpose we consider the expression
\beq
\otimes_{i=1}^N \left(\id_{[i]}+{{\eta}\over{\l-z_i}}H^{(i)}_{\alpha}\right)=
\id_{[N]}+\sum_{i=1}^N {{\eta}\over{\l-z_i}}H^{(i)}_{\alpha}\otimes_{j\neq i}
\left(\id_{[j]}+{{\eta}\over{z_i-z_j}}H^{(j)}_{\alpha}\right).
\eeq
This identity can be proved by noting that both sides have the same limit for 
$\l\rightarrow\infty$ and that the residues at the simple poles $\l=z_i$ are 
identical. If we now consider the order $1/\l$ in the expansion of both sides we obtain
\beq
\sum_{i}H^{(i)}_{\alpha}\otimes_{j\neq i}\id_{[j]}=\sum_{i}H^{(i)}_{\alpha}
\otimes_{j\neq i}\left(\id_{[j]}+{{\eta}\over{z_i-z_j}}H^{(j)}_{\alpha}\right)=\sum_{i}H^{(i)}_{\alpha}\otimes_{j\neq i}
{\tr}^{\alpha}_{(i,j)}{\tilde{{\tr}}}^{\alpha}_{(j,i)}
\eeq
which finishes the proof that the Cartan operators associated with the simple 
roots aquire no dressing, which in turn renders the proof of the commutativity 
of the Cartan operators trivial.\par\noindent
To prove the Serre relation
\beq
\left(ad_{E_{\pm}^{\alpha^{i}}}\right)^{1-A_{ji}}\left(E_{\pm}^{\alpha^{j}}\right)=0
\;\;\;i=1,\ldots,N-1;\;\;i\neq j
\eeq
we have to distinguish 2 cases:
\beq
&1.&\;\;|j-i|=1\;\;\quad\left[E_{\pm}^{\alpha^{i}},\left[E_{\pm}^{\alpha^{i}},
E_{\pm}^{\alpha^{j}}\right]\right]=0\nonumber\\
&2.&\;\;|j-i|>1\;\;\quad\left[E_{\pm}^{\alpha^{i}},E_{\pm}^{\alpha^{j}}\right]=0\label{serre}
\eeq
as all other matrix-elements of the Cartan matrix are zero. (For $sl(n)$ we 
have $A_{ii}=2,\;\;A_{i+1\;i}=A_{i\;i+1}=-1,\;\;A_{ij}=0$ otherwise.)\\
To proceed with the proof we list some useful relations:
\beq
\left(E_{+\alpha}\,{\tr}^{\beta}_{(i,j)}\right)&=&d_{\alpha+1}^{\beta}(i,j)\;
E_{+\alpha}\nonumber\\
\left({\tr}^{\beta}_{(i,j)}\,E_{+\alpha}\right)&=&d_{\alpha}^{\beta}(i,j)\;
E_{+\alpha}\nonumber\\
\left(E_{-\alpha}\,{\tr}^{\beta}_{(i,j)}\right)&=&d_{\alpha}^{\beta}(i,j)\;
E_{-\alpha}\nonumber\\
\left({\tr}^{\beta}_{(i,j)}\,E_{-\alpha}\right)&=&d_{\alpha+1}^{\beta}(i,j)\;
E_{-\alpha}\nonumber\\
\eeq
where $d_{\alpha}^{\beta}(i,j)$ means the $\alpha^{{\rm{th}}}$ element on the 
diagonal of the matrix ${\tr}^{\beta}_{(i,j)}$.\\
We now look at the first case of (\ref{serre}) and show the argument for the positive roots:
\beq
\left[{\tilde{E}}_{\alpha},{\tilde{E}}_{\beta}\right]&=&\sum_{i}
\left[E_{\alpha}^{(i)},E_{\beta}^{(i)}\right]\otimes_{j\neq i}
{\tr}^{\alpha}_{(i,j)}{\tr}^{\beta}_{(i,j)}\nonumber\\
&+&\sum_{i,j} {}^{\prime}{{\eta}\over{z_j-z_i}}E_{\alpha}^{(i)}
\otimes E_{\beta}^{(j)}\otimes_{k\neq i,j} {\tr}^{\alpha}_{(i,k)}
{\tr}^{\beta}_{(j,k)}
\eeq
and thus
\beq
&&\left[{\tilde{E}}_{\alpha},\left[{\tilde{E}}_{\alpha},
{\tilde{E}}_{\beta}\right]\right]\nonumber\\
&=&\sum_{i}\left[E_{\alpha}^{(i)},\left[E_{\alpha}^{(i)},
E_{\beta}^{(i)}\right]\right]\otimes_{j\neq i}{\tr}^{\alpha}_{(i,j)}
{\tr}^{\alpha}_{(i,j)}{\tr}^{\beta}_{(i,j)}\nonumber\\
&+&\sum_{i,j} {}^{\prime}\left(E_{\alpha}^{(i)}
{\tr}^{\alpha}_{(j,i)}{\tr}^{\beta}_{(j,i)}\otimes {\tr}^{\alpha}_{(i,j)}
\left[E_{\alpha}^{(j)},E_{\beta}^{(j)}\right]-{\tr}^{\alpha}_{(j,i)}
{\tr}^{\beta}_{(j,i)}E_{\alpha}^{(i)}\otimes \left[E_{\alpha}^{(j)},
E_{\beta}^{(j)}\right]{\tr}^{\alpha}_{(i,j)}\right)\otimes\nonumber\\
&&\hspace{2em}\otimes_{k\neq i,j} {\tr}^{\alpha}_{(i,k)}
{\tr}^{\alpha}_{(j,k)}{\tr}^{\beta}_{(j,k)}\nonumber\\
&+&\sum_{i,j} {}^{\prime}{{\eta}\over{z_j-z_i}}
\left(E_{\alpha}^{(i)}E_{\alpha}^{(i)}\otimes 
{\tr}^{\alpha}_{(i,j)}E_{\beta}^{(j)}-E_{\alpha}^{(i)}
E_{\alpha}^{(i)}\otimes E_{\beta}^{(j)}{\tr}^{\alpha}_{(i,j)}
\right)\otimes_{k\neq i,j} {\tr}^{\alpha}_{(i,k)}{\tr}^{\alpha}_{(j,k)}
{\tr}^{\beta}_{(j,k)}\nonumber\\
&+&\sum_{i,j} {}^{\prime}{{\eta}\over{z_j-z_i}}\left({\tr}^{\alpha}_{(j,i)}
E_{\alpha}^{(i)}\otimes E_{\alpha}^{(j)}E_{\beta}^{(j)}-E_{\alpha}^{(i)}
{\tr}^{\alpha}_{(j,i)}\otimes E_{\beta}^{(j)}E_{\alpha}^{(j)}\right)
\otimes_{k\neq i,j} {\tr}^{\alpha}_{(i,k)}{\tr}^{\alpha}_{(j,k)}
{\tr}^{\beta}_{(j,k)}\nonumber\\
&+&\sum_{i,j,k} {}^{\prime}{{\eta}\over{z_k-z_j}}\left(E_{\alpha}^{(i)}
{\tr}^{\alpha}_{(j,i)}{\tr}^{\beta}_{(k,i)}\otimes {\tr}^{\alpha}_{(i,j)}
E_{\alpha}^{(j)}\otimes {\tr}^{\alpha}_{(i,k)}E_{\beta}^{(k)}-
{\tr}^{\alpha}_{(j,i)}{\tr}^{\beta}_{(k,i)}E_{\alpha}^{(i)}
\otimes E_{\alpha}^{(j)}{\tr}^{\alpha}_{(i,j)}\otimes E_{\beta}^{(k)}
{\tr}^{\alpha}_{(i,k)}\right)\otimes\nonumber\\
&&\hspace{2em}\otimes_{l\neq i,j,k} {\tr}^{\alpha}_{(i,l)}
{\tr}^{\alpha}_{(j,l)}{\tr}^{\beta}_{(k,l)}\nonumber\\
\eeq
The first term in this sum vanishes due to the Serre relation for the 
undressed operators, the third because $E_{\alpha}E_{\alpha}=0$.\\
The second and the fourth term cancel each other, while the last term 
vanishes for fixed $k$, since the bracket yields 
\beq
{{\eta}\over{z_k-z_j}}\left((1+{{\eta}\over{z_k-z_i}})(1+{{\eta}\over{z_i-z_j}})-
(1+{{\eta}\over{z_j-z_i}})\right)E_{\alpha}^{(i)}\otimes E_{\alpha}^{(j)}\otimes 
E_{\beta}^{(k)}
\eeq
which is antisymmetric under the exchange of $i$ and $j$.\\
The second case  of (\ref{serre}) yields 
\beq
&&\left[{\tilde{E}}_{\alpha},{\tilde{E}}_{\beta}\right]\nonumber\\
&=&\sum_{i}\left[E_{\alpha}^{(i)},E_{\beta}^{(i)}\right]\otimes_{j\neq i}
{\tr}^{\alpha}_{(i,j)}{\tr}^{\beta}_{(i,j)}\nonumber\\
&+&\sum_{i,j} {}^{\prime}\left(E_{\alpha}^{(i)}{\tr}^{\beta}_{(j,i)}\otimes
{\tr}^{\alpha}_{(i,j)}E_{\beta}^{(j)}-{\tr}^{\beta}_{(j,i)}E_{\alpha}^{(i)}
\otimes E_{\beta}^{(j)}{\tr}^{\alpha}_{(i,j)}\right)\otimes_{k\neq i,j} 
{\tr}^{\alpha}_{(i,k)}{\tr}^{\beta}_{(j,k)}
\eeq
where the first term in the sum vanishes due to the assumption for the 
undressed operators and the second term vanishes as the bracket is zero 
for $|\alpha-\beta |>1$.\\
The proof for the $\tilde{E}_{-}^{\alpha^{i}}$ proceeds along the same lines.
\par\noindent
We proceed to give the form of the non--simple roots which can be obtained 
as multiple commutators of simple roots (proof by induction on $\alpha$)
\beqa
\tilde{E}_{i-\alpha\;i}&=&\left[\tilde{E}_{i-\alpha\;i-\alpha+1},\ldots,
\left[\tilde{E}_{i-3\;i-2},\left[\tilde{E}_{i-2\;i-1},\tilde{E}_{i-1\;i}
\right]\right]\ldots\right]\nonumber\\
=&&\sum_{k=1}^{\alpha}\sum_{i_1\neq \ldots\neq i_k}\prod_{\gamma=1}^{k-1}
{{\eta}\over{z_{i_{\gamma}}-z_{i_{\gamma+1}}}}\sum_{\alpha=\beta_0>\beta_1>
\ldots >\beta_k =0}\otimes_{l=1}^k E^{(i_l)}_{i-\beta_{l-1},\,i-\beta_l}\nonumber\\
&&\otimes_{j\neq i_1\ldots i_k}\, \Gamma^{(j)}_{{\underbrace{i_k\ldots i_k}
\atop{\beta_0-\beta_1}}{\underbrace{ i_{k-1}\ldots i_{k-1}}\atop{\beta_1-
\beta_2}}\ldots {\underbrace{i_1\ldots i_1}\atop{\beta_{k-1}-\beta_k}};j;i}
\eeqa
with $\Gamma^{(j)}_{j;i_k\ldots i_k i_{k-1}\ldots i_{k-1}\ldots i_1
\ldots i_1;i}=\mbox{diag}\{1,\ldots ,1,b^{-1}_{i_k j},\ldots,b^{-1}_{i_1 j},
{\underbrace{1, \ldots,1}_{i}}\}_{[j]}$.\\
A similar formula holds for the negative roots
\beqa
\tilde{E}_{i\;i-\alpha}&=&\sum_{k=1}^{\alpha}\sum_{i_1\neq \ldots\neq i_k}
\prod_{\gamma=1}^{k-1}{{\eta}\over{z_{i_{\gamma}}-z_{i_{\gamma+1}}}}
\sum_{\alpha=\beta_0>\beta_1>\ldots >\beta_k =0}\otimes_{l=1}^k 
E^{(i_l)}_{i-\beta_{l},\,i-\beta_{l-1}}\nonumber\\
&&\otimes_{j\neq i_1\ldots i_k}\, \Gamma^{(j)}_{{\underbrace{i_k\ldots i_k}
\atop{\beta_0-\beta_1}}{\underbrace{ i_{k-1}\ldots i_{k-1}}\atop{\beta_1-\beta_2}}
\ldots {\underbrace{i_1\ldots i_1}\atop{\beta_k-1-\beta_k}};j;i}
\eeqa
with $\Gamma^{(j)}_{j;i_k\ldots i_k i_{k-1}\ldots i_{k-1}\ldots i_1\ldots i_1;i}=
\mbox{diag}\{1,\ldots ,1,b^{-1}_{j i_k},\ldots,b^{-1}_{j i_1},{\underbrace{1, \ldots,1}
_{i-1}}\}_{[j]}$.

\section{Appendix B}
In this appendix we discuss further details of the structure of the coefficients 
(cf. (\ref{B_p}))
\beqa
B^{(2)}_p(\lambda_1,\ldots,\lambda_p;z_{i_1},\ldots,z_{i_p})&=&{{\prod_{ij}(\lambda_i-z_j)}
 \over{\prod_{i>j}(\lambda_i-\lambda_j)\prod_{i>j}(z_j-z_i)}}\,\mbox{det} X  \nonumber\\
X_{ij}&=&{1\over{\lambda_i-z_j}}-{1\over{\lambda_i-z_j+\eta}}  .                
\label{det}
\eeqa

This representation, as concise as it is, has the disadvantage that it does not 
reflect the singularity structure in an economic way (the poles in the prefactor 
on the r.h.s. of (\ref{det}) are cancelled by zeroes in the determinant).\\
One may cure the  defect by appropriate manipulations on the determinant in 
(\ref{det}). Subtracting for example the last row of $X$ from all the others, 
extracting a rational factor from the n-th row and proceeding in the same spirit 
with the (n-1)-th row and consecutively other rows one arrives at the equality
\beqa
B^{(2)}_p(\lambda_1,\ldots,\lambda_p;z_{i_1},\ldots,z_{i_p})&=&                        
{{1}\over{\prod_{i<j}(z_j-z_i)}}{{1}\over{\prod_{ij}(\lambda_i-z_j+\eta)}}\,\mbox{det} Y  
\nonumber\\
Y_{\alpha,x}&=&P_{\alpha}(\lambda;\, z_{p_x})\nonumber\\
P_{\alpha}(\lambda;\, z_{p_x})&=&\left\{\prod_{i=0}^{n-\alpha}
(\lambda_{n-i}-z_{p_x}+\eta)-\prod_{i=0}^{n-\alpha}(\lambda_{n-i}-z_{p_x}) 
\right\} \prod_{j=1}^{\alpha-1}(\lambda_j-z_{p_x}+\eta)(\lambda_j-z_{p_x})\nonumber\\
\label{det1}
\eeqa
One should note that the polynomial $P_{\alpha}$ depends on all $\lambda$-variables 
but only on a single $z$-variable. It follows that one can continue to extract 
polynomial factors from the determinant in (\ref{det1}) by subtraction of columns 
from columns. The ensuing differences $P_{\alpha}(\lambda,\,z_{p_x})-P_{\alpha}
(\lambda,\,z_{p_y})$ supply the desired factors $(z_{p_x}-z_{p_y})$ which 
compensate the pole factors ${1\over{\prod_{i>j} (z_i-z_j)}}$ in (\ref{det1}).\\
Unfortunately we have not found a concise closed form for the polynomial multiplying the remaining prefactor ${1\over{\prod_{ij}
(\lambda_i-z_j+\eta)}}$ in the final expression.

\end{appendix}


\begin{thebibliography}{99}\addcontentsline{toc}{section}{References}
\bibitem{ms}Maillet J M, Sanchez de Santos J 1996 
Drinfel'd twists and algebraic Bethe ansatz, q-alg/9612012
\bibitem{drin}Drinfel'd V G 1990 Quasi--Hopf algebras 
{\sl Leningrad Math. J.} {\bf 1} 1419
\bibitem{terras}Terras V 1999 Drinfel'd twists and functional Bethe ansatz 
{\sl Lett. Math. Phys.} {\bf 48} 263-276, math-ph/9902009 
\bibitem{maillet} Kitanine N, Maillet J M, Terras V 1999 Form factors of the 
XXZ Heisenberg spin--1/2 finite chain {\sl Nucl. Phys.} {\bf B 554} 647-678, math-ph/9807020
\bibitem{iz}Izergin A G, Kitanine N, Maillet J M, Terras V 1999 Spontaneous 
magnetization of the XXZ Heisenberg spin-1/2 chain {\sl Nucl. Phys.} 
{\bf B 554} 679-696, solv-int/9812021
\bibitem{qis1}Gohmann F, Korepin V E 1999 Solution of the quantum inverse problem, 
hep-th/9910253
\bibitem{qis2}Maillet J M, Terras V 1999 On the quantum inverse scattering problem,
hep-th/9911030
\bibitem{gaudin}Gaudin M 1976 Diagonalisation d'une classe d'hamiltoniens de spin {\sl Journale de Physique} {\bf 37} 1087
\bibitem{kulresh}Kulish P P, Reshetikin N Y 1981 Generalized Heisenberg ferromagnet 
and the Gross--Neveu model {\sl JETP} {\bf{53}} 108-114
\bibitem{flume}Babujian G M, Flume R 1994 Off-shell Bethe Ansatz equation for 
Gaudin magnets and solutions of Knizhnik-Zamolodchikov equations {\sl Mod. Phys. Lett.}
{\bf A9}  2029-2040, hep-th/9310110
\bibitem{v}De Vega H J 1989 Yang-Baxter algebras, integrable theories and quantum 
groups {\sl Int. J. Mod. Phys.}  {\bf A4}  2371-2463 
\bibitem{bab}Babujian G M 1993 Off-shell Bethe Ansatz equation and N point correlators in SU(2) WZNW theory {\sl Int. J. Mod. Phys.}  {\bf A26} 6981
\bibitem{takht}Takhtadzhyan L A 1983 Quantum inverse scattering method and algebraized 
matrix Bethe--Ansatz {\sl J. Sov. Math.} {\bf 23} 2470-2486
\bibitem{hum}Humphreys J E 1972 
{\sl Introduction to Lie Algebras and Representation Theory} (Berlin: Springer)

\end{thebibliography}
\end{document}